\documentclass{article}
\usepackage{times}
\usepackage{amsfonts}
\usepackage{graphicx}
\begin{document}
\hyphenation{Min-kows-kian}
\noindent
{\Large  THERMALITY OF THE ZERO--POINT LENGTH\\ AND GRAVITATIONAL SELFDUALITY}\\

\vskip1cm
\noindent
{\bf P. Fern\'andez de C\'ordoba}$^{1}$, {\bf J.M. Isidro}$^{2}$ and {\bf Rudranil Roy}$^{3}$\\
Instituto Universitario de Matem\'atica Pura y Aplicada,\\ Universitat Polit\`ecnica de Val\`encia, Valencia 46022, Spain\\
$^{1}${\tt pfernandez@mat.upv.es}, $^{2}${\tt joissan@mat.upv.es},\\
$^{3}${\tt rudranilroy.gravity@gmail.com}
\vskip.5cm
\noindent
\today
\tableofcontents
\vskip.5cm

\noindent
{\bf Abstract}  It has been argued that the existence of a zero--point length is the hallmark of quantum gravity. In this letter we suggest a thermal mechanism whereby this quantum of length arises in flat, Euclidean spacetime $\mathbb{R}^d$. For this we consider the infinite sequence of all flat, Euclidean spacetimes $\mathbb{R}^{d'}$ with $d'\geq d$, and postulate a probability distribution for each $d'$ to occur. The distribution considered is that of a canonical ensemble at temperature $T$, the energy levels those of a 1--dimensional harmonic oscillator. Since both the harmonic energy levels and the spacetime dimensions are evenly spaced, one can identify the canonical distribution of harmonic--oscillator eigenvalues with that of dimensions $d'$. The state describing this statistical ensemble has a mean square deviation in the position operator, that can be interpreted as a quantum of length. Thus placing an oscillator in thermal equilibrium with a bath provides a thermal mechanism whereby a zero--point length is generated. The quantum--gravitational implications of this construction are then discussed. In particular, a model is presented that realises a conjectured duality between a weakly gravitational, strongly quantum system and a weakly quantum, strongly gravitational system.

\section{Introduction}\label{einfuehrung}

It has been argued that first--order quantum--gravity effects manifest themselves through the existence of a quantum of length $L$  \cite{GARAY, KIEFER1, SINGH2}. Under {\it first--order effects}\/ one  understands such as can be observed mesoscopically, without the need for a complete theory of quantum gravity---a theory still under development from a number of different perspectives;  for a sample see {\it e.g.}\/ \cite{KIEFER2} and refs. therein. 

The effects of a zero--point length have been extensively studied in the setting provided by a free, massive, relativistic particle propagating in Euclidean spacetime 
$\mathbb{R}^d$  \cite{PADDY0, PADDY1, PADDY3}.  Let ${\cal S}({\bf x})=\int^{\bf x}{\rm d}s$ be the classical action functional for the particle when $m=1$. Let $G^{d}_{(L=0)}$ be the Feynman propagator for the particle in the {\it absence}\/ of a quantum of length $L$, and let $G^{d}_{(L)}$ denote the same propagator in the {\it presence}\/ of a quantum of length $L$:
\begin{eqnarray}
G^{d}_{(L=0)}({\bf x})&=&\sum_{\rm paths}\exp\left[-m\,{\cal S}({\bf x})\right], \\
\label{1}
G^{d}_{(L)}({\bf x})&=&\sum_{\rm paths}\exp\left\{-m\left[{\cal S}({\bf x})+\frac{L^2}{{\cal S}({\bf x})}\right]\right\}.
\label{sesentaz}
\end{eqnarray}
One can regard the propagator $G^{d}_{(L)}$ as the UV completion of the propagator $G^{d}_{(L=0)}$ because $G^{d}_{(L)}$ is duality invariant, {\it i.e.}\/, invariant under the exchange of ${\cal S}$ and $L^2/{\cal S}$.

Now in ref. \cite{NOI} the decomposition 
\begin{equation}
G^d_{(L)}=\sum_{n=0}^\infty\frac{(-\pi)^{n}}{n!}G^{d+2n}_{(L=0)}
\label{treinta}
\end{equation}
has been established. It expresses the quantum--gravity {\it corrected}\/ propagator in $d$ dimensions as an infinite sum of quantum--gravity {\it free}\/ propagators in all virtual dimensions $d+2n$, with $n\in\mathbb{N}$. This fact seems to cast some doubt on the notion of a sharply defined dimension for quantum spacetimes, at least in the mesoscopic regime.

Let us now Wick rotate Euclidean spacetime $\mathbb{R}^d$ into Minkowski spacetime: $t\rightarrow{\rm i}\tau$.  Let ${\cal G}_M$  (resp. ${\cal G}_R$) denote the Feynman propagator for a real scalar field $\phi$ in the Minkowski vacuum  (resp. Rindler vacuum). Then ${\cal G}_M$ can be thought of as a thermalised version of ${\cal G}_R$, in the sense that \cite{PADDY4, RAJEEV}
\begin{equation}
{\cal G}_M({\rm i}\tau)=\sum_{n=-\infty}^{\infty}{\cal G}_R({\rm i}\tau+2{\rm i}\pi n).
\label{sesentauno}
\end{equation}
Despite the fact that Eq. (\ref{treinta}) refers to a particle while (\ref{sesentauno}) refers to a field, the analogy between them is inspiring: given the thermality of the Rindler frame, it suggests a possible thermal origin for the quantum of length $L$. 

It is one goal of this letter to establish that, indeed, {\it the quantum of length $L$ implementing UV completeness can be interpreted as having a thermal origin}\/. In proving this statement we will achieve a complementary goal. Namely, we will  provide a specific example of the UV/IR duality between {\it the weakly gravitational, strongly quantum regime of a system and the weakly quantum, strongly gravitational regime of a dual system}\/ conjectured in ref. \cite{SINGH1}. 

In order to achieve these two goals we will first identify a canonical ensemble in equilibrium with a thermal bath at temperature $T$, in such a way that it mimics the quantum--gravity corrected, Euclidean spacetime $\mathbb{R}^d_{(L)}$. This ensemble is immediately suggested by the right--hand side of Eq. (\ref{treinta}): an infinite collection of quantum--gravity free, Euclidean spacetimes $\mathbb{R}^{d+2n}_{(L=0)}$, one for each value of $n\in\mathbb{N}$. The following hints provide further clues:\\
{\it i)} associated with a harmonic oscillator there is a natural length scale, namely
\begin{equation}
\lambda_0=\sqrt{\frac{\hbar}{m\omega}};
\label{scala}
\end{equation}
{\it ii)} spacetime dimensions are evenly spaced in the same way as the energy eigenvalues of the 1--dimensional harmonic oscillator;\\
{\it iii)} the sum over virtual dimensions (\ref{treinta}) is such that only those virtual dimensions are summed over that have the same parity as $d$; also this is reminiscent of the well--defined parity of harmonic eigenstates.

These hints suggest that the sought--for canonical ensemble might be given by the infinite collection of excited energy levels $n=d, d+2, d+4, \ldots$ of a 1--dimensional harmonic oscillator. The oscillator groundstate $n=0$ is mapped into the spacetime dimension $d=0$; although the latter is meaningless as a dimension, meaningful physical quantities will be attached to the value $d=0$. This oscillator will be placed in thermal equilibrium with an energy reservoir at temperature $T$. Probabilities will be distributed according to the Boltzmann law: an energy $\varepsilon_n=(n+1/2)\hbar\omega$ will be weighted by $w_n=\exp(-\varepsilon_n/k_BT)$, and the partition function will be given by $Z=\sum_nw_n$. 

The mass $m$ of the harmonic oscillator will be taken to equal that of the particle, and the frequency $\omega$ will be its Compton frequency. The coordinate along which this oscillator moves will be denoted by $q$; the corresponding quantum operator will be $Q$. As usual we will have $H\vert n\rangle = (n+1/2)\hbar\omega\vert n\rangle$, where $H$ is the harmonic Hamiltonian and the $\vert n\rangle$, $n=0,1,\ldots$ are normalised energy eigenstates. This completes our identification of the quantum--mechanical, 1--dimensional harmonic oscillator that is necessary for our construction. What remains is to prove that $L$ actually arises as the thermal average of a certain quantity at a certain temperature; a point that we develop next.

\section{Thermality of the zero--point length}

\subsection{Oscillators as a model for virtual dimensions}\label{zustandsumme}

Let us begin with the canonical partition function $Z_{\rm ho}(\beta)$ for the quantum, 1--dimensional harmonic oscillator in thermal equilibrium with an energy reservoir at temperature $T$:
\begin{equation}
Z_{\rm ho}(\beta)=\sum_{n=0}^{\infty}{\rm e}^{-(n+1/2)\beta\hbar\omega}=\frac{1}{2}\,{\rm csch}\left(\frac{\beta\hbar\omega}{2}\right).
\label{cientocatorce}
\end{equation}
We will also need the partition functions with a defined parity
\begin{equation}
Z_{\rm ho}^{\rm  even/odd}(\beta)=\sum_{n=0\atop {\rm n\, even/odd}}^{\infty}{\rm e}^{-(n+1/2)\beta\hbar\omega}=\frac{1}{2}\,{\rm e}^{\pm\beta\hbar\omega/2}{\rm csch}\left(\beta\hbar\omega\right),
\label{510}
\end{equation}
the positive (resp. negative) sign corresponding to the even (resp. odd) parity.

Moreover, we would like to construct a partition function appropriate to the sum over virtual dimensions (\ref{treinta}). This is readily done: given a value of $d\geq 0$, consider the object $Z_d(\beta)$ defined as
\begin{equation}
Z_d(\beta)=\sum_{n=0}^{\infty}{\rm e}^{-(d+2n+1/2)\beta\hbar\omega}=\frac{1}{2}\,{\rm e}^{-(d-1/2)\beta\hbar\omega}\,{\rm csch}\left(\beta\hbar\omega\right).
\label{doscientosveinte}
\end{equation}
We will refer to the ensemble of oscillator states described by the partition function (\ref{doscientosveinte}) as {\it the truncated oscillator}\/. Truncation means that one sums only over those excited oscillator states $\vert n\rangle$ such that $n\geq d$, and then only over those carrying the same parity as $d$ (as dictated by the sum over virtual dimensions (\ref{treinta})). By contrast we will refer to the ensemble of oscillator states described by the partition function (\ref{cientocatorce}) as {\it the complete oscillator}\/, completeness here meaning that one sums over all states $n\geq 0$, and also regardless of parity.

\subsection{The complete oscillator}

\subsubsection{Thermal density matrices}\label{nullpunkt}

The thermal density matrix for the complete harmonic oscillator is defined by
\begin{equation}
\varrho_{\rm ho}(\beta)=\sum_{n=0}^{\infty}\vert n\rangle {\rm e}^{-(n+1/2)\beta\hbar\omega}\langle n\vert.
\label{matriz}
\end{equation}
We will also need some related sums. One of them is the alternating sum
\begin{equation}
\varrho_{\rm ho}^{\rm alt}(\beta)=\sum_{n=0}^{\infty}(-1)^n\vert n\rangle {\rm e}^{-(n+1/2)\beta\hbar\omega}\langle n\vert,
\label{matrik}
\end{equation}
that one can readily evaluate:
\begin{equation}
\varrho_{\rm ho}^{\rm alt}(\beta)={\rm i}\,\varrho_{\rm ho}\left(\beta+\frac{{\rm i}\pi}{\hbar\omega}\right).
\label{cuatrocientosveintidos}
\end{equation}
Also necessary are the sums over all even/odd states
\begin{equation}
\varrho_{\rm ho}^{\rm even/odd}(\beta)=\sum_{n=0\atop{\rm n\; even/odd}}^{\infty}\vert n\rangle{\rm e}^{-(n+1/2)\beta\hbar\omega}\langle n\vert.
\label{450}
\end{equation}
Now $\varrho_{\rm ho}^{\rm even}(\beta)$ is best evaluated by inserting the projector $(1+(-1)^n)/2$ and then applying Eq. (\ref{cuatrocientosveintidos}), while 
$\varrho_{\rm ho}^{\rm odd}(\beta)=\varrho_{\rm ho}(\beta)-\varrho_{\rm ho}^{\rm even}(\beta)$. We thus arrive at
\begin{equation}
\varrho_{\rm ho}^{\rm even/odd}(\beta)=\frac{1}{2}\,\varrho_{\rm ho}(\beta)\pm\frac{{\rm i}}{2}\,\varrho_{\rm ho}\left(\beta+\frac{{\rm i}\pi}{\hbar\omega}\right),
\label{cuatrocientosveintitres}
\end{equation}
the even (resp. odd) sum corresponding to the plus (resp. minus) sign.

We can now express all the above density matrices in the position representation. Let the matrix elements $\langle q\vert \varrho_{\rm ho}(\beta)\vert q'\rangle$ be denoted by $\varrho_{\rm ho}(q,q';\beta)$. It is known that \cite{WDMR} 
\begin{equation}
\varrho_{\rm ho}(q,q';\beta)
\label{426}
\end{equation}
$$
=\frac{1}{\lambda_0}\frac{1}{\sqrt{2\pi\sinh(\beta\hbar\omega)}}\exp\left\{\frac{-1}{2\lambda_0^2\sinh(\beta\hbar\omega)}\left[(q^2+{q'}^2)\cosh(\beta\hbar\omega)-2qq'\right]\right\}.
$$
Specifically we will need the diagonal matrix element $\langle q\vert \varrho_{\rm ho}(\beta)\vert q\rangle$, also denoted by $\varrho_{\rm ho}(q;\beta)$:
\begin{equation}
\varrho_{\rm ho}(q;\beta)=\frac{1}{\lambda_0}\frac{1}{\sqrt{2\pi\sinh(\beta\hbar\omega)}}\exp\left[-\tanh\left(\frac{\beta\hbar\omega}{2}\right)\frac{q^2}{\lambda_0^2}\right].
\label{470}
\end{equation}
Using Eqs. (\ref{cuatrocientosveintidos}) and (\ref{470}), the diagonal of $\varrho_{\rm ho}^{\rm alt}(\beta)$ turns out to be
\begin{equation}
\varrho_{\rm ho}^{\rm alt}(q;\beta)=\frac{1}{\lambda_0}\frac{1}{\sqrt{2\pi\sinh(\beta\hbar\omega)}}\exp\left[-\coth\left(\frac{\beta\hbar\omega}{2}\right)\frac{q^2}{\lambda_0^2}\right],
\label{471}
\end{equation}
while for the density matrices (\ref{cuatrocientosveintitres}) one finds
\begin{equation}
\varrho_{\rm ho}^{\rm even/odd}(q;\beta)=\frac{1}{2\lambda_0}\frac{1}{\sqrt{2\pi\sinh(\beta\hbar\omega)}}
\label{472}
\end{equation}
$$
\times\left\{\exp\left[-\tanh\left(\frac{\beta\hbar\omega}{2}\right)\frac{q^2}{\lambda_0^2}\right]\pm\exp\left[-\coth\left(\frac{\beta\hbar\omega}{2}\right)\frac{q^2}{\lambda_0^2}\right]\right\},
$$
the positive (resp. negative) sign corresponding to the even (resp. odd) parity.

{}Finally, upon integration over $q\in\mathbb{R}$ we obtain the partition functions 
\begin{equation}
Z_{\rm ho}^{\rm even/odd}(\beta)=\int_{-\infty}^{\infty}{\rm d}q\,\varrho_{\rm ho}^{\rm even/odd}(q;\beta)=\frac{1}{4}\,{\rm csch}\left(\frac{\beta\hbar\omega}{2}\right)\pm\frac{1}{4}\,{\rm sech}\left(\frac{\beta\hbar\omega}{2}\right),
\label{560}
\end{equation}
in perfect agreement with their previous values in Eq. (\ref{510}).

\subsubsection{A thermally induced quantum of length}

It is convenient to reexpress the diagonal matrix elements (\ref{470}) as
\begin{equation}
\varrho_{\rm ho}(q;\beta)=\frac{1}{\lambda_0\sqrt{2\pi\sinh(\beta\hbar\omega)}}\exp\left[-\frac{q^2}{\lambda^2(\beta)}\right],
\label{setenta}
\end{equation}
where the temperature--dependent Gaussian width $\lambda(\beta)$ is the following function:
\begin{equation}
\lambda(\beta)=\lambda_0\sqrt{\coth\left(\frac{\beta\hbar\omega}{2}\right)}.
\label{setentuno}
\end{equation}
Thus $\lambda(\beta)$ is a temperature--dependent length scale analogous to (\ref{scala}), to which it reduces in the zero--temperature limit since $\lim_{\beta\to\infty}\lambda(\beta)=\lambda_0$. Once properly normalised, the Gaussian (\ref{setenta}) gives the probability for finding the system at $Q=q$.  One finds
\begin{equation}
\int_{-\infty}^{\infty}{\rm d}q\,\varrho_{\rm ho}(q;\beta)=Z_{\rm ho}(\beta)
\label{500}
\end{equation}
in nice agreement with the partition function (\ref{cientocatorce}). Thus averages within this thermal ensemble will be computed with respect to the 
normalised distribution $Z^{-1}_{\rm ho}(\beta)\varrho_{\rm ho}(q;\beta)$. 

{}For the position operator $Q$ and its square $Q^2$ we find\footnote{We denote thermal averages by round brackets.}
\begin{equation}
\left(Q\right)_{\rm ho}=\frac{1}{Z_{\rm ho}(\beta)}\int_{-\infty}^{\infty}{\rm d}q\,q\varrho_{\rm ho}(q;\beta)=0
\label{setentados}
\end{equation}
and
\begin{equation}
\left(Q^2\right)_{\rm ho}=\frac{1}{Z_{\rm ho}(\beta)}\int_{-\infty}^{\infty}{\rm d}q\,q^2\varrho_{\rm ho}(q;\beta)=\frac{1}{2}\lambda^2(\beta).
\label{501}
\end{equation}
Then the mean square deviation $\left(\Delta Q\right)_{\rm ho}^2$ reads
\begin{equation}
\left(\Delta Q\right)_{\rm ho}^2=\left( Q^2\right)_{\rm ho}-\left( Q\right)_{\rm ho}^2=\frac{1}{2}\lambda^2(\beta).
\label{setentatres}
\end{equation}
It is meaningful to identify the above mean square deviation with one half the square $L^2$ of the quantum of length for the complete oscillator:
\begin{equation}
\left(\Delta Q\right)_{\rm ho}^2=\frac{1}{2}L^2.
\label{cuatrocientosocho}
\end{equation}
This allows one to solve neatly for the inverse temperature:\footnote{The notation $\coth^{-1}(x)$, sometimes also written ${\rm arccoth}(x)$, stands for the function inverse to $\coth(x)$.}
\begin{equation}
\beta_{\rm ho}=\frac{2}{\hbar\omega}\coth^{-1}\left(\frac{L^2}{\lambda^2_0}\right).
\label{setentacinco}
\end{equation}
We should bear in mind that the averages (\ref{setentados}) and (\ref{501}) are {\it thermal}\/ in nature, because they are computed with respect to the {\it thermal}\/ probability distribution function $Z_{\rm ho}^{-1}(\beta)\varrho_{\rm ho}(q;\beta)$. This notwithstanding, is it instructive to compare them to  the {\it quantum}\/ averages
\begin{equation}
\langle n\vert Q\vert n\rangle=0, \qquad \langle n\vert Q^2\vert n\rangle=\left(n+\frac{1}{2}\right)\lambda_0^2
\label{490}
\end{equation}
corresponding to the harmonic eigenstates $\vert n\rangle$: they match exactly when $n=0$, with the sole replacement of the zero--temperature length scale $\lambda_0$ with its thermal counterpart $\lambda(\beta)$.

Summarising: a zero--point length $L$ defines the temperature of the thermal bath through Eq. (\ref{setentacinco}). Conversely, one can intrepret the latter equation as meaning that {\it a thermal bath induces a quantum of length}\/, as claimed. Instead of string fluctuations as in ref. \cite{FONTANINI}, here we have thermal fluctuations as the origin of the zero--point length.

\subsection{The truncated oscillator}

So far we have only considered the complete oscillator. In the this section we perform a similar analysis for the truncated oscillator.

\subsubsection{Thermal density matrices}

As suggested by the sum over dimensions (\ref{treinta}), here the relevant thermal density matrix to consider is 
\begin{equation}
\varrho_d(\beta)=\sum_{n=0}^{\infty}\vert d+2n\rangle {\rm e}^{-(d+2n+1/2)\beta\hbar\omega}\langle d+2n\vert,
\label{452}
\end{equation}
which we conveniently reexpress as
\begin{equation}
\varrho_d(\beta)=\sum_{n=0\atop{n=d\,{\rm mod}2}}^{\infty}\vert n\rangle {\rm e}^{-(n+1/2)\beta\hbar\omega}\langle n\vert-\sum_{n=0\atop{n=d\,{\rm mod}2}}^{d-1}\vert n\rangle {\rm e}^{-(n+1/2)\beta\hbar\omega}\langle n\vert.
\label{550}
\end{equation}
Above, the parity of the $n$'s summed over is the same as that of $d$.

As before we will denote the diagonal matrix elements $\langle q\vert\varrho_d(\beta)\vert q\rangle$ by $\varrho_{d}(q;\beta)$.  While the integral $\int_{-\infty}^{\infty}{\rm d}q\varrho_d(q;\beta)$ must equal the partition function $Z_d(\beta)$ already known by Eq. (\ref{doscientosveinte}), the integrand $\varrho_d(q;\beta)$ is so far unknown, and must be computed as the diagonal of the density operator (\ref{550}). In the end, $Z_d^{-1}(\beta)\varrho_d(q;\beta)$ will provide us with the normalised probability distribution function necessary to compute thermal averages in this ensemble.  

It turns out that the sought--for distribution function reads (see Eqs. (\ref{573}) and (\ref{580}) of the Appendix)
\begin{equation}
\frac{1}{Z_{d}(\beta)}\varrho^{\rm even/odd}_{d}(q;\beta)=\frac{1}{Z_{d}(\beta)}\left[\varrho_{\rm ho}^{\rm even/odd}(q;\beta)-f_{d}^{\rm even/odd}(q;\beta)\right],
\label{600}
\end{equation}
where ``even/odd" in $\varrho^{\rm even/odd}_{d}(q;\beta)$ refers to the parity of $d$. Above, $Z_d(\beta)$ is given by Eq. (\ref{doscientosveinte}) regardless of parity,  
$\varrho_{\rm ho}^{\rm even/odd}(q;\beta)$ is known by Eq. (\ref{472}), and the functions $f_d^{\rm even/odd}(q;\beta)$ are given by (see Eqs. (\ref{572}) and (\ref{579}) of the Appendix)
\begin{equation}
f_d^{\rm even/odd}(q;\beta)=\sum_{n=0\atop{n\,{\rm even/odd}}}^{d-1} {\rm e}^{-(n+1/2)\beta\hbar\omega}\,\vert\langle n\vert q\rangle\vert^2.
\label{603}
\end{equation}
In (\ref{603}), the parity of the $n$'s summed over is the same as that of $d$.

The thermal probability distribution functions (\ref{600}) differ from their partner (\ref{setenta}) in one important respect, namely, by the terms  $f^{\rm even/odd}_{d}(q;\beta)$. The latter arise from the fact that the lowest--lying energy eigenstate for the truncated oscillator is no longer the (Gaussian) oscillator groundstate $\vert 0\rangle$, but the excited eigenstate $\vert d\rangle$ instead. The finite sums $f_d^{\rm even/odd}(q;\beta)$ are nothing but the thermal probability distributions associated with the oscillator eigenstates lying {\it below}\/ the state $\vert d\rangle$, with due care being taken of parity.

\subsubsection{A thermally induced quantum of length}

We will evaluate the thermal averages
\begin{equation}
\left( Q\right)_{d}^{\rm even/odd}=\frac{1}{Z_d(\beta)}\int_{-\infty}^{\infty}{\rm d}q\,q\varrho_{d}^{\rm even/odd}(q;\beta)
\label{474}
\end{equation}
and
\begin{equation}
\left( Q^2\right)_{d}^{\rm even/odd}=\frac{1}{Z_d(\beta)}\int_{-\infty}^{\infty}{\rm d}q\,q^2\varrho_{d}^{\rm even/odd}(q;\beta).
\label{482}
\end{equation}
As before, a (squared) quantum of length $L^2$ will be defined as (twice) the mean square deviation of the position operator $Q$:
\begin{equation}
\left(\left(\Delta Q\right)_{d}^{\rm even/odd}\right)^2=\left( Q^2\right)_{d}^{\rm even/odd}-\left(\left( Q\right)_{d}^{\rm even/odd}\right)^2=\frac{1}{2}L^2.
\label{475}
\end{equation}
Use of Eqs. (\ref{472}) and (\ref{490}) immediately yields 
\begin{equation}
\left(Q\right)_d^{\rm even/odd}=0.
\label{604}
\end{equation}
However, the evaluation of $\left(Q^2\right)_d^{\rm even/odd}$ is lengthier, and details have been relegated to the Appendix. The final result is Eq. (\ref{625}):
\begin{equation}
\left(\left(\Delta Q\right)_{d}^{\rm even/odd}\right)^2=\left(Q^2\right)_d^{\rm even/odd}=\left[\lambda(2\beta)\right]^2+\lambda_0^2\left(d-\frac{1}{2}\right).
\label{610}
\end{equation}
As was the case for the complete harmonic oscillator, one can interpret the quantum of length $L$ as possessing a thermal origin.

Two features of the above are worth mentioning. First, the temperature--dependent Gaussian width of Eq. (\ref{setentuno}) appears evaluated at $2\beta$ rather than $\beta$. Moreover, there appears a temperature--independent contribution proportional to $(d-1/2)$. These two features arise as consequences of the truncation of the oscillator. The sum over dimensions (\ref{treinta}) starts at the value $d$, and it contains all higher dimensions of the same parity as $d$. This parity requirement amounts to doubling the frequency $\omega$ or, equivalently, the inverse temperature $\beta$.

\section{Gravitational selfduality}\label{dualitaet}

In ref. \cite{SINGH1} it has been conjectured that a strongly quantum, weakly gravitational system must be dual to a weakly quantum, strongly gravitational system.\footnote{A related proposal was put forward in ref. \cite{ISIDRO}.} We claim that an instance of this duality symmetry is provided by the following example.

{}For the complete oscillator of previous sections to implement the weakly gravitational, strongly quantum regime, it suffices to impose two additional requirements:\\
{\it i)} the mass $m$ is very small; \\
{\it ii)} the quantum number $n$ is low enough.

We will construct a dual system that is weakly quantum but strongly gravitational. Consider the quantum system whose (dimensionless) Hamiltonian $\tilde H$ is given by 
\begin{equation}
\tilde H=\left(\frac{H}{\hbar\omega}\right)^{-1},
\label{fckt}
\end{equation}
where $H$ is the harmonic Hamiltonian satisfying $H\vert n\rangle=(n+1/2)\hbar\omega\vert n\rangle$. The (dimensionless) energy levels $\tilde E_n$ of $\tilde H$ are
\begin{equation}
\tilde H\vert n\rangle=\tilde E_n\vert n\rangle, \qquad \tilde E_n=\frac{1}{n+1/2}, \qquad n\in\mathbb{N}.
\label{umkehr}
\end{equation}
Now $n$ was assumed small to guarantee the strongly quantum regime of the initial oscillator; hence the dual system governed by $\tilde H$ implements the weakly quantum behaviour. More precisely: small values of $n$ correspond to eigenvalues of $H$ that are comparable to the oscillator vacuum energy; this may be called the IR regime of the original oscillator. In the dual system governed by $\tilde H$, the same small values of $n$ correspond to high energies (as compared to the rest of the spectrum (\ref{umkehr})); this may be called the UV regime of the dual system. In this sense, {\it the map (\ref{fckt}) between these two dual systems implements UV/IR duality}\/. 

Let the mass of this dual system be $\tilde m$. We need it to be large so gravitational effects will also be large. Thus requiring
\begin{equation}
m\tilde m=M_P^2,
\label{cuiusregio}
\end{equation}
where $M_P$ is the Planck mass, ensures the desired behaviour.

We can further elaborate on the gravitational aspects of the dual system (\ref{umkehr}). The area operator defined as
\begin{equation}
\tilde A=L^2 \tilde H,
\label{cuatrocientoseis}
\end{equation}
where $L$ is the quantum of length, has the quantised area levels
\begin{equation}
\tilde A\vert n\rangle=\tilde A_n \vert n\rangle, \qquad \tilde A_n=\frac{L^2}{n+1/2}, \qquad n\in\mathbb{N}.
\label{cuatrocientosiete}
\end{equation}
We can place this dual system described by the area operator $\tilde A$ in contact with an entropy reservoir at a constant value of the area; ideally this reservoir would be a black hole with horizon area $A_{BH}$ and entropy
\begin{equation}
S_{BH}=kA_{BH}, \qquad k=\frac{k_Bc^3}{4\hbar G}.
\label{doscincuenta}
\end{equation}
Then the dual system can exchange quanta of entropy $k\tilde A_n$ with the entropy reservoir. This is analogous to the oscillator exchanging energy quanta $\hbar\omega$ with the thermal bath. Probabilities are distributed among the different area levels according to the Boltzmann law $\exp\left(-\tilde A_n/A_{BH}\right)$. Furthermore, upon multiplication by $k$ as in Eq. (\ref{doscincuenta}), the area operator $\tilde A$ becomes the entropy operator $\tilde S=k\tilde A$. Then the eigenvalue equation
\begin{equation}
\tilde S\vert n\rangle=\tilde S_n\vert n\rangle, \qquad \tilde S_n= k\tilde A_n
\label{inzz}
\end{equation}
provides an instance of the entropic picture of quantum mechanics first postulated in ref. \cite{ACOSTA}. The duality presented here can also be regarded as an instance of that analysed in ref. \cite{MONDAL}.

\section{Discussion}\label{scuatro}

Gravity has been argued to be selfdual. Indeed gravity escapes the usual pattern of an effective theory (in the Wilsonian sense). Any such theory at low energy holds all the way up to a certain energy scale, beyond which it breaks down and must be replaced by a more fundamental theory. Beginning now with classical gravity at low energy, gravity turns quantum as energy is increased until a certain characteristic scale is reached. Surprisingly, however, gravity becomes once again classical beyond this scale; this is the meaning of selfduality. 

In relation to UV completeness \cite{CROWTHER}, in ref. \cite{SINGH1} it has been argued that a UV/IR duality transformation must exist such that it will map the strongly quantum, weakly gravitational regime of a given system into the strongly gravitational, weakly quantum regime of a dual system. In this article we have presented an explicit example of two systems exhibiting this UV/IR duality property.

Our starting point was Eq. (\ref{treinta}), which states that quantum--gravity properties are conferred upon the $d$--dimensional propagator $G^d_{(L)}$ by an infinite sum of propagators $G^{d+2n}_{(L=0)}$ in virtual dimensions; the latter are all classical in the sense that they all carry $L=0$. This has led us to interpret the zero--point length, the hallmark of quantum gravity, as a thermal phenomenon. Novel attempts at unification \cite{FINSTER1, FINSTER2, SINGH3, SINGH4} also take the quantum of length into account. Moreover, thermality should not surprise us given that thermodynamics is central to a number of modern approaches to gravity \cite{JACOBSON, PADDY5, VERLINDE1, VERLINDE2}. Reinterpreted from a thermal point of view, the duality invariance of the quantum--mechanical propagator (\ref{treinta}) is perfectly natural: the infinite sum is over all propagators {\it with a given parity}\/. Definite parity implies duality invariance; undefined parity does not.

We have found that, while zero--temperature dimensions are sharply defined, thermal dimensions become averages over a statistical ensemble of virtual dimensions of classical spacetimes. Transitions between different virtual dimensions are induced by thermal fluctuations. Thus virtual dimensions fluctuate, as they should in any quantum theory. We have modelled virtual dimensions on the quantum number of a 1--dimensional harmonic oscillator. However we have not altered the notion of a smooth spacetime manifold in any other way: as stated in section \ref{einfuehrung}, our analysis centers around the mesoscopics of quantum gravity, where all quantum effects can be ascribed to a nonvanishing quantum of length. The theory developed here is not of quantum {\it spacetime}\/, but of the {\it dimension}\/ of quantum spacetime in the mesoscopic regime. The microscopic constituents, or {\it atoms}\/, of the dimension of a quantum--gravity  corrected spacetime $\mathbb{R}^d_{(L)}$ are the virtual dimensions $d+2n$ of all quantum--gravity free spacetimes $\mathbb{R}^{d+2n}_{(L=0)}$ for $n\in\mathbb{N}$. A model for these virtual dimensions has been provided by the energy levels of a truncated oscillator.

Altogether, we can view the thermality of the zero--point length as yet another argument in favour of an atomistic, or granular, nature of quantum spacetime.

\vskip1cm
\noindent
{\bf Acknowledgments} Work funded by FEDER/MCIN under grant PID2021-128676OB-I00.

\section{Appendix}

\subsection{Computation of $\varrho_d^{\rm even/odd}(q;\beta)$}

Here we derive Eq. (\ref{600}), analysing the cases $d=2k$ and $d=2k+1$ separately.

{\it i)} $d=2k$. Then the density matrix (\ref{550}) becomes
\begin{equation}
\varrho_{d=2k}(\beta)=\sum_{n=0\atop{n\,{\rm even}}}^{\infty}\vert n\rangle {\rm e}^{-(n+1/2)\beta\hbar\omega}\langle n\vert-\sum_{n=0\atop{n\,{\rm even}}}^{d-1}\vert n\rangle {\rm e}^{-(n+1/2)\beta\hbar\omega}\langle n\vert.
\label{551}
\end{equation}
The first sum equals the density operator $\varrho_{\rm ho}^{\rm even}(\beta)$ of Eq. (\ref{450}), the corresponding diagonal being given in Eq. (\ref{472}). Therefore the diagonal entries of (\ref{551}) read
\begin{equation}
\varrho_{d=2k}(q;\beta)=\varrho_{\rm ho}^{\rm even}(q;\beta)-\sum_{n=0\atop{n\,{\rm even}}}^{d-1} {\rm e}^{-(n+1/2)\beta\hbar\omega}\,\vert\langle n\vert q\rangle\vert^2.
\label{552}
\end{equation}
Finally defining
\begin{equation}
f_{d=2k}^{\rm even}(q;\beta)=\sum_{n=0\atop{n\,{\rm even}}}^{d-1} {\rm e}^{-(n+1/2)\beta\hbar\omega}\,\vert\langle n\vert q\rangle\vert^2,
\label{572}
\end{equation}
the sought--for diagonal matrix element is
\begin{equation}
\varrho_{d=2k}(q;\beta)=\varrho_{\rm ho}^{\rm even}(q;\beta)-f^{\rm even}_{d=2k}(q;\beta),
\label{573}
\end{equation}
where the right--hand side is explicitly known by Eqs. (\ref{472}) and (\ref{572}). As a consistency check, integration over $q\in\mathbb{R}$ should produce the partition function $Z_{d=2k}(\beta)$. Indeed:
\begin{equation}
Z_{d=2k}(\beta)=Z_{\rm ho}^{\rm even}(\beta)-\sum_{n=0\atop{n\,{\rm even}}}^{d-1} {\rm e}^{-(n+1/2)\beta\hbar\omega},
\label{570}
\end{equation}
on account of Eq. (\ref{560}) and of the $\vert n\rangle$ being normalised eigenstates. Now the finite geometric sum can be readily computed,
\begin{equation}
\sum_{n=0\atop n\,{\rm even}}^{d-1} {\rm e}^{-(n+1/2)\beta\hbar\omega}=\frac{1}{2}\left[{\rm e}^{\beta\hbar\omega/2}-{\rm e}^{-(d-1/2)\beta\hbar\omega}\right]{\rm csch}(\beta\hbar\omega),
\label{532}
\end{equation}
and Eqs. (\ref{560}) and (\ref{532}) yield the partition function $Z_{d=2k}(\beta)$
\begin{equation}
Z_{d=2k}(\beta)=\frac{1}{2}\,{\rm e}^{-(d-1/2)\beta\hbar\omega}{\rm csch}(\beta\hbar\omega),
\label{571}
\end{equation}
in happy agreement with its previous evaluation in Eq. (\ref{doscientosveinte}). 

{\it ii)} $d=2k+1$. Here the density matrix (\ref{550}) becomes
\begin{equation}
\varrho_{d=2k+1}(\beta)=\sum_{n=0\atop{n\,{\rm odd}}}^{\infty}\vert n\rangle {\rm e}^{-(n+1/2)\beta\hbar\omega}\langle n\vert-\sum_{n=0\atop{n\,{\rm odd}}}^{d-1}\vert n\rangle {\rm e}^{-(n+1/2)\beta\hbar\omega}\langle n\vert.
\label{574}
\end{equation}
The first sum equals the density operator $\varrho_{\rm ho}^{\rm odd}(\beta)$, the corresponding diagonal being given in Eq. (\ref{472}). Therefore the diagonal entries of (\ref{574}) read
\begin{equation}
\varrho_{d=2k+1}(q;\beta)=\varrho_{\rm ho}^{\rm odd}(q;\beta)-\sum_{n=0\atop{n\,{\rm odd}}}^{d-1} {\rm e}^{-(n+1/2)\beta\hbar\omega}\,\vert\langle n\vert q\rangle\vert^2,
\label{575}
\end{equation}
so defining
\begin{equation}
f^{\rm odd}_{d=2k+1}(q;\beta)=\sum_{n=0\atop{n\,{\rm odd}}}^{d-1} {\rm e}^{-(n+1/2)\beta\hbar\omega}\,\vert\langle n\vert q\rangle\vert^2,
\label{579}
\end{equation}
the sought--for diagonal matrix element will be
\begin{equation}
\varrho_{d=2k+1}(q;\beta)=\varrho_{\rm ho}^{\rm odd}(q;\beta)-f^{\rm odd}_{d=2k+1}(q;\beta),
\label{580}
\end{equation}
where the right--hand side is known by Eqs. (\ref{472}) and (\ref{579}). As a check, integration over $q\in\mathbb{R}$ should produce the partition function $Z_{d=2k+1}(\beta)$. This is indeed the case:
\begin{equation}
Z_{d=2k+1}(\beta)=Z_{\rm ho}^{\rm odd}(\beta)-\sum_{n=0\atop{n\,{\rm odd}}}^{d-1} {\rm e}^{-(n+1/2)\beta\hbar\omega},
\label{576}
\end{equation}
again on account of Eq. (\ref{560}) and of the normalisation of the $\vert n\rangle$. Moreover,
\begin{equation}
\sum_{n=0\atop n\,{\rm odd}}^{d-1} {\rm e}^{-(n+1/2)\beta\hbar\omega}=\frac{1}{2}\left[{\rm e}^{-\beta\hbar\omega/2}-{\rm e}^{-(d-1/2)\beta\hbar\omega}\right]{\rm csch}(\beta\hbar\omega).
\label{577}
\end{equation}
Thus Eqs. (\ref{560}) and (\ref{577}) yield the partition function $Z_{d=2k+1}(\beta)$:
\begin{equation}
Z_{d=2k+1}(\beta)=\frac{1}{2}\,{\rm e}^{-(d-1/2)\beta\hbar\omega}{\rm csch}(\beta\hbar\omega),
\label{578}
\end{equation}
again in beautiful agreement with Eq. (\ref{doscientosveinte}).

\subsection{Computation of $\left(Q^2\right)_d^{\rm even/odd}$}

By Eq. (\ref{600}),
\begin{equation}
\int_{-\infty}^{\infty}{\rm d}q\,q^2\varrho_d^{\rm even/odd}(q;\beta)=\int_{-\infty}^{\infty}{\rm d}q\,q^2\varrho_{\rm ho}^{\rm even/odd}(q;\beta)
-\int_{-\infty}^{\infty}{\rm d}q\,q^2f_d^{\rm even/odd}(q;\beta).
\label{611}
\end{equation}
Terms to evaluate are
\begin{equation}
\int_{-\infty}^{\infty}{\rm d}q\,q^2\,\varrho_{\rm ho}^{\rm even/odd}(q;\beta)
\label{605}
\end{equation}
$$
=\frac{\lambda_0^2}{8}\left[\coth\left(\frac{\beta\hbar\omega}{2}\right){\rm csch}\left(\frac{\beta\hbar\omega}{2}\right)\pm\tanh\left(\frac{\beta\hbar\omega}{2}\right){\rm sech}\left(\frac{\beta\hbar\omega}{2}\right)\right]
$$
where Eq. (\ref{472}) has been applied,  and
\begin{equation}
\int_{-\infty}^{\infty}{\rm d}q\,q^2f_d^{\rm even/odd}(q;\beta)=\lambda_0^2\sum_{n=0\atop{\rm n\;even/odd}}^{d-1}\left(n+\frac{1}{2}\right){\rm e}^{-(n+1/2)\beta\hbar\omega}
\label{606}
\end{equation}
after using Eqs. (\ref{490}), (\ref{603}). The finite geometric sums are straightforward to evaluate as the derivatives, with respect to $\beta\hbar\omega$, of the sums (\ref{532}) and (\ref{577}). Then adding together Eqs. (\ref{606}) and (\ref{605}) produces, after some algebra, 
\begin{equation}
\int_{-\infty}^{\infty}{\rm d}q\;q^2\varrho_d^{\rm even/odd}(q;\beta)
=\frac{\lambda_0^2}{4}\,{\rm e}^{-(d-1/2)\beta\hbar\omega} {\rm csch}(\beta\hbar\omega) \left[2 d + 2\, {\rm coth}(\beta\hbar\omega) - 1\right].
\label{630}
\end{equation}
{}Finally using Eq. (\ref{setentuno}) we conclude
$$
\left(Q^2\right)_d^{\rm even/odd}=Z_d(\beta)^{-1}\int_{-\infty}^{\infty}{\rm d}q\;q^2\varrho_d^{\rm even/odd}(q;\beta)
$$
\begin{equation}
=\frac{\lambda_0^2}{2} \left[2 d -1+ 2\, {\rm coth}(\beta\hbar\omega) \right]=\left[\lambda(2\beta)\right]^2+\lambda_0^2\left(d-\frac{1}{2}\right).
\label{625}
\end{equation}

\end{document}